\documentstyle[12pt]{article}
\newfont{\Mb}{msbm10}

\newcommand{\Od}[2]{{{{\rm d} #1}\over{{\rm d} #2}}}

\begin{document}
\setcounter{equation}{0}
\setcounter{figure}{0}
\setcounter{table}{0}

\title{Prelle-Singer Approach and Liouvillian Solutions: An Important Result}
\author{
L.G.S. Duarte\thanks{
Universidade do Estado do Rio de Janeiro,
Instituto de F\'{\i}sica, Departamento de F\'{\i}sica Te\'orica,
R. S\~ao Francisco Xavier, 524, Maracan\~a, CEP 20550--013,
Rio de Janeiro, RJ, Brazil. E-mail: lduarte@dft.if.uerj.br},
S.E.S. Duarte\thanks{
idem. E-mail: sduarte@dft.if.uerj.br}, 
and L.A.C.P. da Mota\thanks{
idem. E-mail: damota@dft.if.uerj.br}
}
\maketitle
\abstract{In \cite{firsTHEOps1}, we present a method to tackle first order ordinary differential equations whose solutions contain Liouvillian functions (LFOODEs), many of them missed by the usual PS-approach. Here, we demonstrate an important result concerning the generality of that method.}
\newpage
\section{Introduction}
The problem of solving ordinary differential equations (ODEs) has led,
over the years, to a wide range of different methods for their solution.
Along with the many techniques for calculating tricky integrals, these
often occupy a large part of the mathematics syllabuses of university
courses in applied mathematics round the world. 

The overwhelming majority of these methods 
are based on classification of the DE into types for which a method
of solution is known, which has resulted in a gamut of methods that
deal with specific classes of DEs. This scene changed somewhat at the
end of the 19th century when Sophus Lie developed a general method to
solve (or at least reduce the order of) ordinary differential equations
(ODEs) given their symmetry transformations ~\cite{step,bluman,olver}. Lie's method
is very powerful and highly general, but first requires that we find
the symmetries of the differential equation, which may not be easy to do.
Search methods have been developed~\cite{nosso,nosso2} to extract
the symmetries of a given ODE, however these methods are heuristic and
cannot guarantee that, if symmetries exist, they will be found.

A big step forward in constructing an
algorithm for solving first order ODEs (FOODEs) analytically was taken in a
seminal paper by Prelle and Singer (PS)~\cite{PS} on autonomous systems of
ODEs. Prelle and Singer's problem is equivalent to asking when a FOODE
of the form $y'=M(x,y)/N(x,y)$, with $M$ and $N$ polynomials in their
arguments, has an elementary solution (a solution which can be written
in terms of a combination of polynomials, logarithms, exponentials
and radicals). Prelle and Singer were not exactly able to construct an
algorithm for solving their problem, since they were not able to define
a degree bound for the polynomials which might enter into the
solution. Though this is important from a theoretical point of view, any
pratical use of the PS method will have a degree bound imposed by the 
time necessary to perform the actual calculation needed to handle the
ever-more complex equations. With this in mind it is possible to say
that Prelle and Singer's original method is almost an algorithm, awaiting
a theoretical degree bound to turn it algorithmic. In \cite{firsTHEOps1}, we present 
a method which is an extension to the Prelle-Singer procedure allowing for the solution of some LFOODEs. This method is very effective in solving a class of LFOODEs missed by the usual Prelle-Singer procedure. It deppended though, 
for being general, on a result which we did not prove at the time. Here, we present such a proof thus guaranteeing that the method is generally applicable.
The paper is organized as follows: in section~\ref{PSreview}, we present a short
theoretical introduction to the PS approach and show why the Prelle-Singer aproach misses some LFOODEs; in section~\ref{theorem}, we
demonstrate a theorem concerning the general structure of the integrating factor for LFOODEs of a certain type; we
finally present our conclusions.
\section{The PS-Method and FOODEs with Liouvillian Functions in the Solutions}
\label{PSreview}
\subsection{The usual PS-Method}
\label{usual}
Despite its usefulness in solving FOODEs, the Prelle-Singer procedure is
not very well known outside mathematical circles, and so we present
a brief overview of the main ideas of the procedure.
Consider the class of FOODEs which can be written as
\begin{equation}
\label{FOODE}
y' = {\frac{dy}{dx}} = {\frac{M(x,y)}{N(x,y)}}
\end{equation}
where $M(x,y)$ and $N(x,y)$ are polynomials with coefficients in the complex 
field $\it C$.
If $R$ is a integrating factor of~(\ref{FOODE}), it is possible to write
$\partial_x (RN) + \partial_y (RM) = 0$, leading to 
$N\, \partial_x R + R\,\partial_y N + M\, \partial_y R + R\, \partial_y M = 0.$ Thus, we finally obtain:
\begin{equation}
\label{eq_PS}
{\frac{D[R]}{R}} = - \left( \partial_x N + \partial_y M\right),
\end{equation}
where $D \equiv N \partial_x + M \partial_y.$
In \cite{PS}, Prelle and Singer showed that, if the solution of (\ref{FOODE}) is written in terms of elementary functions, then $R$ must be of the form $R = \prod_i f^{n_i}_i$ where $f_i$ are irreducible polynomials and $n_i$ are non-zero rational numbers. Using this result in (\ref{eq_PS}), we have
\begin{eqnarray}
\label{ratio}
{\frac{D[R]}{R}} & = & {\frac{D[\prod_{i} f^{n_i}_i]}{\prod_i f^{n_k}_k}} =
{\frac{\sum_i f^{n_i-1}_i n_i D[f_i] \prod_{j \ne i}
f_j^{n_j}}{\prod_k f^{n_k}_k}} \nonumber \\[3mm] 
& = & \sum_i {\frac{f^{n_i-1}_i n_i D[f_i]}{f_i^{n_i}}} =
\sum_i {\frac{n_iD[f_i]}{f_i}}.
\end{eqnarray}
From~(\ref{eq_PS}), plus the fact that $M$ and $N$ are polynomials, 
we conclude that ${D[R]}/{R}$ is a polynomial. One can then prove from~(\ref{ratio}) that $f_i | D[f_i]$ \cite{PS}.
We now have a criterion for choosing the possible $f_i$ (build all
the possible divisors of $D[f_i]$) and, by using~(\ref{eq_PS})
and~(\ref{ratio}), we have
\begin{equation}
\label{eq_ni}
\sum_i {\frac{n_iD[f_i]}{f_i}} = - \left( \partial_x N + \partial_y M\right).
\end{equation}
If we manage to solve~(\ref{eq_ni}) and thereby find $n_i$,
we know the integrating factor for the FOODE and the problem is
reduced to a quadrature.
\subsection{FOODEs with Liouvillian Functions in the Solutions}
\label{liou}
In this section, we will explain why the PS-method fails to solve some FOODEs which solutions present Liouvillian functions.
As we have already mentioned, the usual PS-method garantees to find a solution
for a FOODE if it is expressible in terms of elementary functions. However, the
method can also solve some FOODEs with non-elementary solutions. Why and why
not for all the FOODEs with non-elementary solutions? 
Consider the following two examples:
\begin{equation}
\label{Kamke211}
\Od{y}{x}=\frac {3\,{x}^{2}y^{2}+{x}
^{3}+1}{4 \left (x+1\right )\left ({x}^{2}-x+1\right )y}
\end{equation}
and
\begin{equation}
\label{Kamke21}
\Od{y}{x}= y^{2}+y x+x-1,
\end{equation}
equation I.18 of the standard testing 
ground for ODE solvers by Kamke~\cite{kamke}.
These equations present (respectively) general solutions given by:
\begin{equation}
\label{solKamke211}
y^{2}-\sqrt {x+1}\sqrt {{x}^{2}-x+1} \left(
1/2\,\int \!{
\frac {1}{\sqrt {x+1}\sqrt {{x}^{2}-x+1}}}{dx}-2{\it \_C1} \right) =0,
\end{equation}
and
\begin{equation}
\label{solKamke21}
y =-{\frac {2\,{\it \_C1}+\sqrt {-1}\sqrt {\pi }{e^{-2}}\sqrt {2}{
\it erf}(1/2\,\sqrt {-1}\sqrt {2}x-\sqrt {-1}\sqrt {2})-2\,{e^{1/2\,x
\left (x-4\right )}}}{2\,{\it \_C1}+\sqrt {-1}\sqrt {\pi }{e^{-2}}
\sqrt {2}{\it erf}(1/2\,\sqrt {-1}\sqrt {2}x-\sqrt {-1}\sqrt {2})}}
.
\end{equation}
We note that both solutions are not expressible in terms of elementary
functions. But, for FOODE \ref{Kamke211}, the standard PS-method can find
the solution (eq. \ref{solKamke211}). The same is not true for FOODE
\ref{Kamke21}, what is happening? 
This can be best understood if we have
a look on the integrating factors for those FOODEs, which are respectively:
\begin{equation}
\label{R211}
R = \left ({x}^{3}+1\right )^{-3/2},
\end{equation}
and
\begin{equation}
\label{R21}
R = {\frac {{e^{{x}^{2}/2-2\,x}}}{\left (y+1\right )^{2}}}.
\end{equation}
Since the standard PS procedure constructs integrating factor candidates from
polynomials in the variables ($x,y$), one can see that, since the integrating factor on eq. \ref{R21} presents the exponential
$e^{{x}^{2}/2-2\,x}$, it will never be found by the PS-method. 
So, can we understand something about the general structure of the FOODE that
will allow us, eventually, to solve a class of these latter equations?
In \cite{firsTHEOps1}, we have demonstrated that, if (\ref{FOODE}) is a LFOODE, its integrating factor will be of the form:
\begin{equation} 
\label{generalformofR}
R = e^{r_0(x,y)} \prod_{i=1}^{n} p_i(x,y)^{c_i}.
\end{equation} 
where $r_0$ is a rational function on $(x,y)$, the $p_i$'s are irreducible polynomials on $(x,y)$ and the $c_i$'s are constants. 
Applying this into eq. (\ref{eq_PS}), we get:
\begin{equation}
\label{eqadendum}
D[r_0(x,y)] + \sum_i {\frac{c_iD[p_i]}{p_i}} = - \left( \partial_x N + \partial_y M \right),
\end{equation}
With this result, we could build a procedure to tackle certain classes of LFOODEs (missed by the usual PS-method) \cite{firsTHEOps1}. In the bulk of the procedure, we use the hypothesis (based on our experience in dealing with those FOODEs) that $p_i$ are ``eigenpolynomials'' of the $D$ operator. Our procedure proved to be successful in solving a class of such FOODEs and we actualy could not find a single counterexample where our hypothesis (and consequently, our method) is not applicable. In the next section, we are going to prove that such a hypothesis is true and, consequently, our procedure is, indeed, generaly applicable.
\section{A Theorem concerning the structure of the integrating factor for a LFOODE}
\label{theorem}
As we have already mentioned, in \cite{firsTHEOps1} we demonstrated that, for a LFOODE, the integrating factor has the general form given by eq.(\ref{generalformofR}). Also there, we have prsented a method of solving LFOODEs that, with the (by the time of writing \cite{firsTHEOps1}) hypothesis concerning the $p_i$'s (mentioned above), can be broadly divided into two steps:
\begin{enumerate}
\item {\bf The determination of the $p_i$'s:} Since they are ``eigenpolynomials'' of the $D$ operator, is done in the same fashion as it is done in the usual PS-method (see section \ref{usual} above).
\item {\bf The determination of $r_0(x,y)$:} In step 1, we get the $p_i$'s and, from eq.(\ref{eqadendum}), we solve for $r_0(x,y)$, see \cite{firsTHEOps1}.
\end{enumerate}
In this section, we are going to demonstrate that the above mentioned hypothesis is actualy a reality, therefore completing our analysis of the general form for the integrating factor for LFOODEs and, furthermore, assuring (by this demonstration) that the method presented on \cite{firsTHEOps1} is generally applicable.
\bigskip
\bigskip
{\bf Theorem 1:} If we have a LFOODE of the form $dy/dx=M(x,y)/N(x,y)$, where $M(x,y)$
and $N(x,y)$ are polynomials in $(x,y)$, with integrating factor $R$ given by $R = e^{r_0(x,y)} \prod_{i=1}^{n} p_i(x,y)^{c_i}$, where $r_0$ is rational function on $(x,y)$, $p_i$ are irreducible polynomials on $(x,y)$ and $c_i$ are constants, then $D[r_0]$ is a polynomial on $(x,y)$, where $D \equiv N \partial_x + M \partial_y.$
\bigskip
{\bf Proof:} Since $r_0$ is a rational function, we can write (\ref{eqadendum}) as:
\begin{equation}
\label{EQ_DR}
D\left[\frac{P(x,y)}{Q(x,y)}\right] + \sum_i c_i \frac{D[p_i]}{p_i} = - \left( \partial_x N + \partial_y M\right),
\end{equation}
where $P$ and $Q$ are polynomials on $(x,y)$ with no common factors. Writing $\sum_i c_i \frac{D[p_i]}{p_i}$ as a single quotient, we get:

\begin{equation}
\label{EQ_DR2}
D\left[\frac{P}{Q}\right] + \frac{\sum_j c_j (\prod_{i,i\neq j}p_i) D[p_j]}{\prod_i p_i} = - \left( \partial_x N + \partial_y M\right).
\end{equation}

Expanding $D\left[\frac{P}{Q}\right]$ and multiplying both sides of (\ref{EQ_DR2}) by $\prod_i p_i$, we can write:
\begin{equation}
\label{EQ_DR3}
\prod_i p_i \,\,\frac{Q\,D[P] - P\,D[Q]}{Q^2} + \sum_j c_j (\prod_{i,i\neq j}p_i) D[p_j] = - \left( \partial_x N + \partial_y M\right) (\prod_i p_i ).
\end{equation}
Since $\sum_j c_j (\prod_{i,i\neq j}p_i) D[p_j]$ is a polynomial and so is the right-hand side of (\ref{EQ_DR3}), we can conclude that 
\begin{equation}
\label{p1}
\prod_i p_i \,\,\frac{Q\,D[P] - P\,D[Q]}{Q^2}
\end{equation}
is a polynomial.
Since $p_i$ are irreducible polynomials, $\prod_i p_i $ can not divide $Q^2$. So, we have two main situations:
\begin{enumerate}
\item $\prod_i p_i$ and $Q$ have no common factors. 
\item $\prod_i p_i$ and $Q$ have common factors.
\end{enumerate}
Let us consider the first situation: Since, by assumption, $\prod_i p_i$ does not have any common factor with $Q$ and that $\prod_i p_i \,\,\frac{Q\,D[P] - P\,D[Q]}{Q^2}$ is a polynomial, we must have that
\begin{equation}
\label{p2}
\frac{Q\,D[P] - P\,D[Q]}{Q^2}
\end{equation}
is itself a polynomial. 
Since $(Q\,D[P] - P\,D[Q])/Q^2 = D[r_0]
$, the proof is complete for this case.
The situation for the second case is a little more involved. First, let us consider that, in $\prod_i p_i$, $i$ runs from $1$ to $n$. With that in mind, let us stablish some notation:
Representing the common factor of $Q$ and $\prod_{i=1}^n p_i$ as:
\begin{equation}
{\cal I} = \prod_{i=1}^{n_1} p_i,
\end{equation}
and the terms in $\prod_{i=1}^n p_i$ not present in $Q$ as:
\begin{equation}
{\pi} = \prod_{i=n_1+1}^{n} {p}_i,
\end{equation}
we can write:
\begin{equation}
\label{pitheta}
\prod_{i=1}^{n} p_i = \pi\,\,{\cal I},\,\,\,\,\,\,\,\,\,\,\,\,Q = \theta\,\,{\cal I} 
\end{equation}
where $\theta = \prod_{i=1}^{n_{\theta}} {q}_i^{m_i}$ and $m_i$ are positive integers.
Re-writing (\ref{p1}) with this notation and expanding, we obtain:
\begin{eqnarray}
\label{p2}
\prod_i p_i \,\,\frac{Q\,D[P] - P\,D[Q]}{Q^2} & \rightarrow & ( \pi\,\,{\cal I})\,\,\,\,\frac{Q\,D[P] - P\,D[Q]}{Q\,\theta\,\,{\cal I}} \nonumber\\
\rightarrow\,\, \pi\,\,\,\frac{Q\,D[P] - P\,D[Q]}{Q\,\theta}&\rightarrow&\pi\,\,\frac{D[P]}{\theta} - \pi\,P\, \frac{D[Q]}{Q\,\theta}
\end{eqnarray}
Remembering that (\ref{p2}) is a polynomial (see (\ref{p1})), if we multiply (\ref{p2}) by $\theta$ (itself a polynomial), we get that
\begin{equation}
\pi\,D[P] - \pi\,P\, \frac{D[Q]}{Q}
\end{equation}
is a polynomial. Therefore, since $\pi\,D[P]$ is a polynomial, we finaly may conclude that
\begin{equation}
\pi\,P\, \frac{D[Q]}{Q}
\end{equation}
is a polynomial. From the fact that neither $\pi$ nor $P$ have factors in common with $Q$, we can assure that $D[Q]/Q$ is a polynomial.
Let us consider the situation $2$ (therefore coompleting the possible cases): Reminding the reader that, by (\ref{pitheta}), $Q = \theta\,\,{\cal I} = \left(\prod_{i=1}^{n_{\theta}} {q}_i^{m_i}\right)\,\left(\prod_{i=1}^{n_1} p_i\right)
$ and, using a similar line of reasonig that we used for $R$ arround eq.(\ref{ratio}), we can conclude that $q_i | D[q_i], i=1..n_{\theta}$ and $p_i | D[p_i], i=1..n_{1}$. 
Writing eq.(\ref{eqadendum}) as:
\begin{equation}
\label{eqadendum2}
D[r_0(x,y)] + \sum_{i=1}^{n_1} {\frac{c_iD[p_i]}{p_i}} + \sum_{i=n_1+1}^{n} {\frac{c_iD[p_i]}{p_i}} = - \left( \partial_x N + \partial_y M \right),
\end{equation}
and, since $p_i | D[p_i], i=1..n_{1}$, we have that $\sum_{i=1}^{n_1} {\frac{c_iD[p_i]}{p_i}}$ is a polynomial. Therefore
\begin{equation}
\label{eqadendum3}
D[r_0] + \sum_{i=n_1+1}^{n} {\frac{c_iD[p_i]}{p_i}} 
\end{equation}
is a polynomial. We can write (\ref{eqadendum3}) as:
\begin{equation}
\label{eqadendum33}
D\left[\frac{P}{Q}\right] + \frac{\sum_{j} c_j (\prod_{i,i\neq j}p_i) D[p_j]}{\prod_{i} p_i},\,\,\,\,i,j=n_1+1\cdots n.
\end{equation}
Multiplying (\ref{eqadendum33}) by $\prod_{i} p_i$, we get:
\begin{equation}
\label{EQ_DR33}
\prod_i p_i \,\,\frac{Q\,D[P] - P\,D[Q]}{Q^2} + \sum_j c_j (\prod_{i,i\neq j}p_i) D[p_j],\,\,\,\,i,j=n_1+1\cdots n 
\end{equation}
which is also a polynomial. Since $\sum_j c_j (\prod_{i,i\neq j}p_i) D[p_j]$ is itself a polynomial, we can safely say that:
\begin{equation}
\prod_{i=n_1+1}^n p_i \,\,\,\,\frac{Q\,D[P] - P\,D[Q]}{Q^2} 
\end{equation}
is a polynomial. Since $\prod_{i=n_1+1}^n p_i$ has no common factor with $Q$, we finally conclude that $\frac{Q\,D[P] - P\,D[Q]}{Q^2}=D[r_0]$ is polynomial.
Thus, we have demonstrated that $D[r_0]$ is always polynomial.
\bigskip
\bigskip
\bigskip
{\bf Corollary 1:} If $R = e^{r_0(x,y)} \prod_{i=1}^{n} p_i(x,y)^{c_i}$ (where $r_0$ is a rational function on $(x,y)$, the $p_i$'s are irreducible polynomials on $(x,y)$ and the $c_i$'s are constants) is the integrating factor for the LFOODE $y' = M/N$, where $M$, $N$ are polynomials in $(x,y)$, then $p_i | D[p_i]$.
The result above is a direct consequence of eq.\,(\ref{eqadendum}).
\section{Conclusion}
In \cite{firsTHEOps1}, we have developed a method to deal with a class of LFOODEs. This method proved to be very effective and we could not find (within the class under study) any counterexample where it was not applicable. Nevertheless, in order to prove its genetrality, the result presented on this paper was lacking. For that reason only, our present result is justified. But, apart from that, it is worth mentioning that our result completes the determination of the general form for the integrating factor for any LFOODE of the type given by eq. (\ref{FOODE}). Now we can safelly say that the integrating factor for any LFOODE has the form:
\begin{equation} 
\label{generalformofRfinal}
R = e^{r_0(x,y)} \prod_{i=1}^{n} p_i(x,y)^{c_i}.
\end{equation} 
where $D[r_0]$ is a rational function on $(x,y)$, the $p_i$'s are ``eigenpolynomials'' of the $D$ operator and the $c_i$'s are constants. 

\end{document}